# Magnon polaron formed by selectively coupled coherent magnon and phonon modes of a surface patterned ferromagnet

F. Godejohann[1*], A. V. Scherbakov[1,2*], S. M. Kukhtaruk[1,3], A. N. Poddubny[2], D. D. Yaremkevich[1], M. Wang[4], A. Nadzeyka[5], D. R. Yakovlev[1,2], A. W. Rushforth[4], A. V. Akimov[4], and M. Bayer[1,2]

[1]*Experimentelle Physik 2, Technische Universität Dortmund, 44227 Dortmund, Germany.*

[2]*Ioffe Institute, 194021 St. Petersburg, Russia.*

[3]*Department of Theoretical Physics, V.E. Lashkaryov Institute of Semiconductor Physics, 03028 Kyiv, Ukraine.*

[4]*School of Physics and Astronomy, University of Nottingham, Nottingham NG7 2RD, United Kingdom.*

[5]*Raith GmbH, 44263 Dortmund, Germany*

**ABSTRACT**

Strong coupling between two quanta of different excitations leads to the formation of a hybridized state which paves a way for exploiting new degrees of freedom to control phenomena with high efficiency and precision. A magnon polaron is the hybridized state of a phonon and a magnon, the elementary quanta of lattice vibrations and spin waves in a magnetically-ordered material. A magnon polaron can be formed at the intersection of the magnon and phonon dispersions, where their frequencies coincide. The observation of magnon polarons in the time domain has remained extremely challenging because the weak interaction of magnons and phonons and their short lifetimes jeopardize the strong coupling required for the formation of a hybridized state. Here, we overcome these limitations by spatial matching of magnons and phonons in a metallic ferromagnet with a nanoscale periodic surface pattern. The spatial overlap of the selected phonon and magnon modes formed in the periodic ferromagnetic structure results in a high coupling strength which, in combination with their long lifetimes allows us to find clear evidence of an optically excited magnon polaron. We show that the symmetries of the localized magnon and phonon states play a crucial role in the magnon polaron formation and its manifestation in the optically excited magnetic transients.

*Corresponding author.
felix.godejohann@tu-dortmund.de; alexey.shcherbakov@tu-dortmund.de.

**INTRODUCTION**

Magnons are collective spin excitations in magnetically-ordered materials. Nowadays the manipulation of coherent high-frequency magnons on the nanoscale is one of the most prospective concepts for information technologies, also in the quantum regime. In this respect, the hybridization of magnons with phonons has been considered as a powerful method for spin control [1-6]. The magnon-phonon hybridization phenomenon in bulk materials is well understood [7,8] and has been realized experimentally in the MHz frequency range [9,10]. However, on the nanoscale, where magnon and phonon frequencies reach the GHz and sub-THz frequency ranges, magnon-phonon hybridized states, referred to further as magnon polarons, are not yet comprehensively understood or utilized.

Within the last decade, the magnon-phonon interaction has been actively studied in experiments with high-frequency coherent magnons and phonons [11-20]. For instance, the excitation of coherent magnons has been realized by broadband coherent phonon wave packets [11,13,19,20], localized monochromatic phonons [14,15,17,18] and propagating surface acoustic waves [12,16]. However, these experiments demonstrate only the one-way process of transferring energy from phonons to magnons with no evidence of strong coupling, characterized by a reversible energy exchange between them, and formation of a hybridized state.

Direct evidence of the formation of magnon polarons would be the avoided crossing of the magnon and phonon dispersion curves at their intersection [7,8]. A spectral splitting between the two branches arising from hybridization clearly indicates strong coupling of two underlying excitations, like in the case of two strongly coupled harmonic oscillators [21,22]. The experimental observation of the avoided crossing for magnon polarons is possible when the energy splitting at the magnon-phonon resonance exceeds the spectral widths of the interacting modes. Only recently, experiments with Ni nanomagnets [23] have demonstrated that these conditions can be approached by means of quantization of the magnon and phonon spectra by a three-dimensional confinement. Quantitatively, the hybridization is characterized by the cooperativity $C = \kappa^2/\gamma_{\text{ph}}\gamma_{\text{M}}$ where $\kappa$ is the coupling strength, i.e. the energy exchange rate between phonons and magnons, and $\gamma_{\text{ph}}$ and $\gamma_{\text{M}}$ are their individual damping rates. High cooperativity $C \gg 1$ ensures deterministic manipulation of the coupled states with periodic conversion of magnons into phonons and vice versa with high fidelity, but is extremely challenging to realize in practice due to the weak magnon-phonon interaction and broad magnon and phonon spectra with respective quick dephasing of these excitations. The latter factor is crucial for ferromagnetic metals.

Here, we demonstrate an original approach for achieving pronounced hybridization of magnons and phonons with high cooperativity in a metallic ferromagnet. By surface patterning of a ferromagnetic film, we achieve magnon and phonon modes with significantly long lifetimes. Spatial matching of their wave distributions determines the coupling strength and selects particular modes in the phonon and magnon spectra for hybridization. This allows us to solve the main problem of quick dephasing of the interacting excitations, and we achieve $C \approx 8$ sufficient for pronounced magnon polaron formation. The suggested approach is verified in time-resolved experiments revealing the avoided crossing between the magnetic polaron branches in the transient magnon evolution that is excited by a femtosecond optical pulse. We demonstrate and discuss the scenarios in which coherent magnon polarons are generated and detected depending on the spatial symmetry of the phonons and magnons in the nanostructure and on the initial and boundary conditions of the coupled system. Our theoretical analysis based on coupled oscillators illustrates the physics behind the experimental observations.

**GALFENOL NANOGRATING AND EXPERIMENTAL TECHNIQUES**

To verify the suggested approach, we use a magnetostrictive alloy of iron and gallium ($Fe_{0.81}Ga_{0.19}$), known as Galfenol. This metal possesses both enhanced magnon-phonon interaction [24] and well-defined magnon resonances [25,26]. A film of $Fe_{0.81}Ga_{0.19}$ of 105-nm thickness was epitaxially grown on a (001)-GaAs substrate after 10 periods of a GaAs/AlAs (59 nm/71 nm) superlattice. The Galfenol film was capped by a 3-nm thick Cr layer to prevent oxidation. The film was patterned into a shallow one-dimensional nanograting (NG) of 25×25-$\mu m^2$ size, which is illustrated in Fig. 1(a). Parallel grooves in the sample surface along the [010] crystallographic axis of the GaAs substrate were milled using a focused beam of Ga ions (Raith VELION FIB-SEM). The grooves have depth *a*=7 nm and width *w*=100 nm, which equals their separation; the respective NG lateral period is *d*=200 nm. The sample was fixed by a silver paste to a massive copper plate which served as a heat sink and was located between the poles of an electromagnetic coil. The measurements were performed at ambient conditions.

The experimental technique shown in Fig. 1(b) is based on conventional magneto-optical pump-probe spectroscopy. The pump-probe scheme was realized by means of two mode-locked Erbium-doped ring fiber lasers (TOPTICA FemtoFiber Ultra 1050 and FemtoFiber Ultra 780). The lasers generate pulses of 150-fs duration with a repetition rate of 80 MHz at wavelengths of 1046 nm (pump pulses) and 780 nm (probe pulses). The energy density in the focused pump spot of 5-$\mu$m diameter was 12 mJ/cm$^2$. The energy density in the 1-$\mu$m-diameter spot of the linearly polarized probe pulse at the NG surface was 1 mJ/cm$^2$. The temporal resolution was achieved by means of an asynchronous optical sampling (ASOPS) technique [27]. The pump and probe oscillators were locked with a frequency offset of 800 Hz. In combination with the 80-MHz repetition rate, it allowed measurement of the time-resolved signals in a time window of 12.5 ns with time resolution limited by the probe pulse duration. To measure the time evolution of the NG magnetization, we use a detection scheme that monitors the polar Kerr rotation (KR) of the probe's polarization plane, $\Psi(t)$, where $t$ is the time delay between the probe and pump pulses [28]. Transient KR was detected by means of a differential scheme based on a Wollaston prism and a balanced optical receiver with 10-MHz bandwidth. By measuring the intensity of the reflected probe pulse, $\Delta I(t)$, we monitor the coherent photoelastic response of the NG [29]. The modulation of the probe pulse intensity was measured by a single photodiode with no polarization optics on the optical path of the probe beam reflected from the sample. The main measurements were carried out in the geometry with excitation of the NG by the pump pulses through the GaAs substrate, which is transparent at the pump pulse wavelength. The probe measurements for characterization of the interacting phonon and magnon modes were performed at the front-side of the NG. An external magnetic field, **B**, applied in the NG plane was used to adjust the magnon spectrum.

For obtaining the main parameters of the magnon and phonon modes as well as their hybridized state, the transient KR and reflectivity signals were analyzed by fast Fourier transform (FFT) and by fitting. For the FFT analysis we used rectangular envelope windows with width of 10 ns. This allows us to obtain the frequencies and spectral widths of the main harmonics contributing to the transient signal with a resolution of 0.1 GHz. For fitting of the transient signals we use the following function: $y(t) = \sum A_j e^{-t/\tau_j} \sin(2\pi f_j t - \varphi_j)$, where $A_j, \tau_j, f_j,$ and $\varphi_j$ are the amplitude, lifetime, frequency and phase of the $j^{th}$ harmonic contributing to the measured transient signal. A summation by three harmonics was enough for fitting of any experimental signal with high accuracy. The spectral power density of this function obtained by Fourier transform consists of Lorentzian peaks centered at $f_j$. Their spectral width determined as half width at half maximum (HWHM) $\gamma_j = 1/2\pi\tau_j$. The estimated errors of $f_j$ and $\gamma_j$ are ±0.002 GHz and ±0.02 GHz, respectively.

**PHONON AND MAGNON MODES IN THE NANOGRATING AND THE EXPERIMENTAL CONCEPT**

The phonon spectrum generated by the pump pulse in the NG includes several phonon modes localized in the near-surface region. These modes are excited due to the strong absorption (20-nm penetration depth) of the pump pulse in Galfenol. The calculated spatial profiles of the atom displacements corresponding to the localized modes are shown in Fig. 1(c). Both modes are standing pseudo-surface acoustic waves [30], which possess two displacement components along the *x*- and *z*-axis. Therefore, they are characterized by three components of the dynamical strain: $\eta_{xx}, \eta_{zz}, \eta_{xz}$, the frequencies and phases of which remain preserved within the respective lifetimes of the modes. The mode shown in the lower panel of Fig. 1(b) is a Rayleigh-like standing wave with dominant displacement along the *z*-axis, i.e. perpendicular to the NG plane. We refer to this mode as a quasi-transverse acoustic (QTA) mode. It is important to mention that there is a counterpart of the QTA mode, which has different symmetry properties and cannot be excited optically. This localized mode is almost degenerate with the QTA mode, but as we show below, can play a crucial role in the interactions with magnons.

Another mode, shown in the top panel of Fig. 1(c) is often referred to as a surface skimmed longitudinal mode [17]. It has a predominant in-plane displacement and we refer to this mode as a quasi-longitudinal acoustic (QLA) mode. The calculated frequencies [31] of the QTA and QLA modes are 13.1 and 15.3 GHz, respectively. The QTA and QLA modes are excited simultaneously and are expected to have similar amplitudes and lifetimes. However, due to the specific polarization and spatial distribution, the QLA mode provides much less contribution to the reflectivity signal [31]. This can be seen in Fig. 1(d), which shows $\Delta I(t)$ and its fast Fourier transform (FFT) measured at $B$=250 mT applied along the NG grooves when the interaction of the magnon and phonon modes of the NG is fully suppressed [32]. The lower spectral line at $f_{QTA}$=13.0 GHz has a large amplitude, while the upper, QLA mode, is invisible. The lifetime of the QTA mode obtained from the fit of the transient reflectivity signal is 5.3 ns, which agrees with the spectral width (HWHM) $\gamma_{QTA} = 0.03$ GHz of the respective line in the power density spectrum obtained by FFT. The corresponding *Q*-factor of the QTA phonon mode exceeds 200.

Now we consider the magnon spectrum of the NG. Because the studied grating is formed by shallow grooves, the depths of which are much smaller than the grating period and the ferromagnetic film thickness, the magnon spectrum is close to that of an unpatterned film [33]. The transient KR signal measured for a plain Galfenol film (outside the NG) is shown in Fig. 1(e). The fast decaying oscillations reflect the magnetization precession excited by the femtosecond optical pulse, which induces ultrafast changes of the magnetic anisotropy [26, 34, 35]. The magnon spectrum of a plain film consists of several magnon modes quantized along the *z*-axis [26,34]. The high-order modes possess short lifetimes, which manifests as fast decay of the precession within several hundred picoseconds after optical excitation. However, at *t*>1 ns the fundamental magnon (FM) mode with the lowest frequency $f_{FM}$=18.2 GHz at *B*=200 mT contributes solely to the KR signal. In the whole range of the magnetic field used in the experiment, it has a considerably long lifetime $\tau_{FM}$=0.95 ns with respective HWHM of the power density spectrum $\gamma_{FM}$=0.17 GHz. In the NG, the magnon modes possess additional spatial modulation along the *x*-axis with the NG period *d* due to the periodically modulated demagnetizing field (shape anisotropy) given by the NG spatial profile [33]. Despite the modulation along the *x*-direction, the FM mode remains the most pronounced in the magnon spectrum of the NG and possesses the same spectral width and a similar dependence of the spectral position on *B* as in the plain Galfenol film [33]. The main object of our study is the interaction of this FM mode with the two localized phonon modes, QTA and QLA.

The idea of our experiments is demonstrated schematically in Fig. 1(f), which shows the predicted magnetic field dependences of the frequencies for uncoupled phonon and FM modes. For phonons, the frequencies are independent of *B* and are shown as horizontal lines. For the FM mode, the dependence on *B* is linear in a certain range of magnetic field. The magnon-phonon hybridization phenomena are expected at the crossing points. Experimentally, we measure the transient KR signals at various *B*. By analyzing the FFTs of the measured transient signals, we obtain information about the FM spectrum paying special attention to the resonance conditions.

**AVOIDED CROSSING AT THE MAGNON-PHONON RESONANCE**

Figure 2 summarizes our main experimental observations. The color contour map in Fig. 2(a) shows the field dependence of the spectral amplitude density of the transient KR signal, measured at **B** applied along the NG diagonal when the interaction between the magnon and phonon modes has maximal strength [32]. The pivotal result is the well-resolved avoided crossing observed at the magnon-phonon resonance located at $B$=110 mT. The normal mode splitting at the nominal intersection of the FM and QTA modes, $\Delta$, obtained by a fit to the transient KR signal and from its FFT, is quantified as $\Delta$ = 0.4GHz. The presence of the avoided crossing and the value of $\Delta$ are demonstrated also in the inset of Fig. 2(a), which shows the field dependence of the frequencies of the two spectral peaks in the measured FFT spectra around 13.0 GHz. Figure 2(b) shows the transient KR signals and their spectra obtained for off-resonant ($B$=30 mT) and resonant ($B$=110 mT) conditions. For off-resonance the spectrum consists of a broad magnon band including a peak at $f_{FM}$, that corresponds to the FM mode, and two intense narrow peaks at the frequencies of the QTA and QLA phonon modes. Their existence in combination with high amplitudes is due to the driving of higher-order magnon modes in the NG by the localized QTA and QLA phonon modes [32]. For on-resonance when the frequency of the FM mode, $f_{FM}$, coincides with the frequency of the QTA phonon mode, $f_{QTA}$, the normal mode splitting around $f=f_{FM}=f_{QTA}$ is clearly observed [see also the zoomed section in Fig. 2(c)], which uniquely indicates the avoided crossing. This gives direct evidence of the magnon polaron, i.e. the optically excited, hybridized magnon-phonon state formed at the FM-QTA resonance.

The resonance of the FM mode with the QLA mode is evidently seen as the bright red spot in Fig. 2(a) at $B$=140 mT when $f_{FM}=f_{QLA}$. The transient KR signal and its FFT measured at this resonance are shown in Fig. 3(a). The strong increase of the spectral amplitude at the QLA mode frequency is explicitly seen from comparison of the FFT spectra obtained in and out of the FM-QLA resonance and shown in the left panel of Fig. 3(b). This increase is the result of resonant phonon driving of magnetization precession [12,15-18,23], which clearly indicates magnon-phonon interaction. However, no avoided crossing is observed at the intersection of the FM and QLA modes.

Summarizing the main experimental results, we clearly observe magnon polaron formation for the FM-QTA resonance through the normal mode splitting, but no avoided crossing is detected for the FM-QLA resonance. Another important difference between the two resonances is the strong increase of the KR signal due to phonon driving, which is observed only for the FM-QLA resonance.

**SYMMETRIES OF THE MODES AND COUPLING SELECTIVITY**

To understand the differences in the manifestations of the two magnon-phonon resonances, we analyze the magnon-phonon interaction in the NG in more detail. Our analysis is based upon the approach developed in Ref. [36]. It has been shown that the coupling strength of interacting magnon and phonon modes can be determined by the spatial overlap of the dynamical magnetization, $\delta \boldsymbol{m}$, of a magnon mode and the strain components of a phonon mode. In this case, the interacting magnon and phonon modes can be considered as two coupled oscillators [36]. Due to the in-plane orientation of the external magnetic field the *z*-component of the steady-state magnetization can be assumed to be zero. In this case, for modeling the magnon-phonon interaction we may consider only two strain components: $\eta_{xx}$ and $\eta_{xz}$ [32]. The coupling strength, $\kappa \approx \Delta/2$, for the magnon and phonon modes at resonance is defined by two overlap integrals:

$$\kappa = \beta_1 \int \tilde{\eta}_{xx} \widetilde{\delta m}_x dV + \beta_2 \int \tilde{\eta}_{xz} \widetilde{\delta m}_z dV, \qquad (1)$$

where $\widetilde{\delta m}_{x,z}$ and $\tilde{\eta}_{xx,xz}$ are the projections of the dynamical magnetization of the magnon mode and the strain components of the phonon mode, respectively, normalized in such a way that $\int \widetilde{\delta m}_{x,z}^2 dV = \int \tilde{\eta}_{xx,xz}^2 dV = 1$ ($dV$ is a dimensionless unit volume element). The coefficients $\beta_1$ and $\beta_2$, which have dimension of frequency, are defined by the material parameters including the magneto-elastic coefficients, the saturation magnetization $M_s$, the mass density of the media, as well as the external magnetic field orientation and the resonant frequency [37].

To evaluate Eq. (1) we consider the spatial distributions of the phonon modes localized in the NG. As mentioned earlier, there is a counterpart of the QTA mode referred to as QTA*. The spatial distributions of the strain components are shown in the bottom of Fig. 4(a). The QTA and QTA* modes are split due to Bragg reflections and interferences in the spatially periodic NG [38], but their splitting does not exceed 0.1 GHz. Thus, the QTA and QTA* modes may be considered as degenerate. The calculated HWHM of both modes, which is determined by escape to the substrate, is much smaller than the measured value of 0.03 GHz for the QTA mode, which is determined by imperfections of the NG. Thus, we may assume $\gamma_{QTA*}=\gamma_{QTA}$. The only difference between the QTA and QTA* mode is the symmetry along the *x*-axis. We refer to the QTA mode as *symmetric* due to the symmetry of the strain components $\tilde{\eta}_{xx}$ and $\tilde{\eta}_{zz}$ relative to the center of the groove of the NG. The QTA* mode is referred to as *antisymmetric*. Only the symmetric QTA mode is excited by the pump pulse [31]. The QTA* mode cannot be excited by the pump pulse due to the antisymmetric nature of $\tilde{\eta}_{zz}$, and it also cannot be optically detected [31]. However, the shear strain component $\tilde{\eta}_{xz}$, has the opposite symmetry which will be important for the interaction with magnons: it is antisymmetric for QTA and symmetric for QTA* [see Fig. 4(a)].

Next, we consider the magnon spatial distribution, assuming mixed boundary conditions for the magnetization in the studied structure along the *z*-axis: pinning at the (Fe,Ga)/GaAs interface and free precession at the patterned surface of the NG [39]. Then, the spatial distribution of the normalized dynamical magnetization, $\widetilde{\delta m}$, for the FM mode can be written as

$$\widetilde{\delta m_x} = \widetilde{\delta m_z} = A \cos\left(\frac{2\pi}{d}x\right)\sin\left(\frac{\pi}{2h}z\right) \qquad (2)$$

where z=0 corresponds to the (Fe,Ga)/GaAs interface. The spatial distribution of $\widetilde{\delta m}$ is shown in the center of Fig. 4(a). The calculations based on Eqs. (1) and (2) show that the overlaps of the FM mode distribution with both $\tilde{\eta}_{xx}$ and $\tilde{\eta}_{xz}$ for the QTA mode are negligible due to their poor spatial match: the overlap integrals for both strain components are less than $10^{-3}$. In contrast, the FM mode closely matches the $\tilde{\eta}_{xz}$ of the antisymmetric QTA* mode, which leads to the coupling strength $\kappa_{\text{QTA}^*} \approx 0.77\beta_2$. Thus, the coupling strength of the FM mode for the QTA* mode is three orders of magnitude stronger than for the QTA mode in the NG. We conclude that the experimentally observed hybridization corresponds to the coupling of the FM with the antisymmetric QTA* mode. The magnon polaron is excited via the excitation of the FM mode by the optical pump pulse and detected by the probe pulse in the KR signal, also via the FM mode. The phonon QTA* mode is not excited optically and, therefore, driving of the FM mode by the QTA* mode does not take place, which agrees with the experimental observation for the lower magnon-phonon resonance at $B$ = 110 mT. This explains the experimental observation of the avoided crossing effect without resonant driving by phonons, as demonstrated in Fig. 2. The value of the measured spectral splitting $\Delta = 0.4$ GHz at the resonance is close to the calculated value of $\Delta = 0.28$ GHz for the frequency splitting of a pure transverse wave propagating in bulk Galfenol along the [100] crystallographic direction at 45 degrees to the applied magnetic field [31].

A similar analysis for the QLA mode, for which the spatial distribution is demonstrated in the upper part of Fig. 4(a), shows that the overlap of the FM mode with the optically excited QLA mode is nonzero ($\kappa_{\text{QLA}} \approx \beta_1$) and driving of the FM mode by the QLA mode should take place. This agrees with the experimental results in Fig. 3(b) where the increase of the KR signal in resonance is clearly seen. The absence of a normal mode splitting in this case could be related to the smaller value of $\kappa_{\text{QLA}}$, for which the normal mode splitting becomes affected by the energy transfer from the optically-excited coherent phonons to magnons [23].

**MODEL OF COUPLED OSCILLATORS AND COOPERATIVITY OF THE MAGNON-PHONON COUPLING**

For better understanding of the interaction of the phonon and magnon modes in the NG and its experimental manifestations, we consider a model of three interacting oscillators using the following system of equations:

$$\frac{1}{2\pi}\frac{da_j}{dt} + \gamma_j a_j + if_j a_j - i\sum_l \text{K}_{jl} a_l = A_j \delta(t), \qquad (3)$$

where $a_j$ are the complex amplitudes of the respective modes ($j$=QTA*, QLA, or FM) and $A_j$ is the amplitude of their excitation. The coupling tensor $\widehat{\text{K}}$ has the form

$$\widehat{\text{K}} = \begin{pmatrix} 0 & 0 & \kappa_{\text{QTA}^*} \\ 0 & 0 & \kappa_{\text{QLA}} \\ \kappa_{\text{QTA}^*} & \kappa_{\text{QLA}} & 0 \end{pmatrix}, \qquad (4)$$

where $\kappa_{QTA^*}$ and $\kappa_{QLA}$ are the coupling strengths of the respective phonon mode and the FM mode. At the resonance of the magnon mode with the certain phonon mode the solutions for the resonance frequency read:

$$f_{R,l} = f_l - i\frac{\gamma_{FM}+\gamma_l}{2} \pm \sqrt{\kappa_l^2 - \left(\frac{\gamma_{FM}-\gamma_l}{2}\right)^2}, \qquad (5)$$

where $l$ stands for QTA* or QLA, $f_l$ is the frequency of the uncoupled QTA* or QLA modes. The real and imaginary part of $f_{R,l}$ gives the frequency and damping of the corresponding hybridized magnon-phonon mode. In the case of strong coupling, when $\kappa_l > |\frac{\gamma_{FM}-\gamma_l}{2}|$, the hybridized state possesess the normal mode splitting and the decay rate for the split peaks is given by the arithmetic average $\gamma_R = \frac{\gamma_{FM}+\gamma_l}{2}$. In the case of weak (or moderate) coupling, where $\kappa_l \leq |\frac{\gamma_{FM}-\gamma_l}{2}|$, the coupling affects the decay rate and the decay rates for the coupled state is given by

$$\gamma_{R\pm} = \frac{\gamma_{FM}+\gamma_l}{2} \mp \sqrt{\left(\frac{\gamma_{FM}-\gamma_n}{2}\right)^2 - \kappa_l^2}. \qquad (6)$$

The suggested model allows us to extract the parameters of the magnon-phonon coupling at the FM-QTA* and FM-QLA resonances from the transient signals and their FFTs. At the FM-QTA* resonance the measured spectral splitting $\Delta = 2\kappa_{QTA^*} = 0.4$ GHz, exceeds both the decay rates for the FM mode, $\gamma_{FM} = 0.17$ GHz, and the QTA* mode, $\gamma_{QTA^*} = 0.03$ GHz, respectively. Thus, the estimated cooperativity of the magnon-phonon coupling at the intersection of the FM and QTA* modes: $C_{QTA^*} = \kappa_{QTA^*}^2/(\gamma_{QTA^*}\gamma_{FM}) = 7.8$. This value exceeds threshold $C = 1$ by almost an order of magnitude, ensuring the strong coupling regime and the hybridization of magnons and phonons for on-resonance conditions. An additional confirmation of the achieved strong coupling is the spectral width of the split peaks in Fig. 2(c) of $\gamma_R = 0.10$ GHz, which corresponds to the arithmetic average of $\gamma_{FM}$ and $\gamma_{QTA^*}$.

The coupling strength of the FM and QLA modes, $\kappa_{QLA}$, cannot be extracted directly from the resonant transient signal due to the absence of the avoided crossing effect. However, it can be estimated from the analysis of the spectral widths of the interacting modes in and out of resonance [40]. With the assumption of weak FM-QLA coupling Eq. (6) gives $\kappa_{QLA}^2 = (\gamma_R - \gamma_{QLA})(\gamma_{FM} - \gamma_R)$, where $\gamma_R = 0.10$ GHz is the spectral width of the line at the resonance $f = f_{FM} = f_{QLA}$ shown in Fig. 3(a). The spectral width of the QLA mode, $\gamma_{QLA}$, was not measured directly due to its extremely weak contribution to $\Delta I(t)$, but we may assume that it is determined also by the NG imperfections and $\gamma_{QLA} = \gamma_{QTA} = 0.03$ GHz. Thus, the coupling strength of the FM and QLA modes is $\kappa_{QLA} = 0.07$ GHz, which provides a cooperativity $C_{QLA} = 1$. For such "moderate coupling", the formation of the magnon polaron may still take place, but the avoided crossing would be masked by the energy transfer from the phonon mode [31].

The difference between manifestations of the experimentally observed regimes of the magnon-phonon coupling is clearly seen in the color map shown in Figs. 4(b) and 4(c), which has been obtained by analytical solution of Eq. (3) with the parameters given above. The chosen excitation amplitudes are $A_{\mathrm{QLA}} = 10$; $A_{\mathrm{FM}} = 1$; $A_{\mathrm{QTA}^*} = 0$ corresponding to the experimentally realized situation when the QTA* mode is not excited and the energy transferred from the optical pump pulse to the QLA mode is 100 times higher than the energy injected into the FM mode. The calculated magnetic field dependence shows strong similarities with the experimental color map in Fig. 2, so that the model can be considered as prototypical for the involved physics. The main features of the magnon spectra in the calculated color map of Fig. 4(b), i.e. the avoided crossing and the driving, are in good agreement with the experimentally measured magnon spectra shown in Fig. 2(a). Indeed, the avoided crossing effect is observed for the lower FM-QTA* resonance mode while for the upper FM-QLA resonance the splitting of the hybridized state is masked by strong driving of the FM mode by resonant phonons. It is worth mentioning, that if the modes coupled with *C* = 1 are excited with equal initial amplitudes an avoided crossing is well resolved [31].

**MAGNON POLARON IN THE REFLECTIVITY SIGNAL**

The magnon-phonon coupling and the formation of a magnon polaron should be manifested also in the field dependence of the transient reflectivity signal. The spectral amplitude of the QTA mode in $\Delta I(t)$ does not depend strongly on *B*, which agrees with our model where the symmetric QTA mode does not interact with the FM mode. We might expect an effect of the interaction between the symmetric QTA mode with the optically inactive antisymmetric counterpart of the FM magnon mode, which is also present in the spatially periodic NG. Actually, the spatial distribution of the antisymmetric FM mode along the *z*-axis is altered by the demagnetizing field of the NG and differs significantly from the distribution of the symmetric FM mode resulting in a much smaller overlap integral and the absence of strong coupling with the QTA mode. That is why the avoided crossing for the QTA mode is not observed.

In contrast to the QTA mode, the contribution of the QLA mode to $\Delta I(t)$ demonstrates a strong dependence on *B.* Out of resonance the contribution of the QLA mode to the transient reflectivity is extremely weak, which is apparently due to an inefficient photo-elastic effect. It increases significantly at the FM-QLA resonance at *B*=140 mT as demonstrated in the right panel in Fig. 3(b) and in Fig. 3(c). Such an increase can be due to the magnon-phonon interaction, which distorts the polarization of the phonon mode similarly to how it happens in the bulk [41] and is predicted for surface acoustic waves [42]. The in-plane shear component, $\eta_{xy}$, makes the QLA mode visible in the photo-elastic effect [31]. In this case, the reflectivity signal, Δ*I*(*t*), for the QLA mode increases at the resonance field (*B*=140 mT), which is observed in our experiment. However, the influence of quadratic magneto-optical Kerr effects [43] on the reflectivity signal at the resonance conditions, when the precession amplitude is large, cannot be fully excluded.

**CONCLUSIONS**

To conclude, we have studied experimentally the properties of the optically excited magnon-phonon hybrid excitations in a ferromagnetic nanograting. By changing the external magnetic field, we are able to realize resonance conditions for a magnon mode with two localized phonon modes with different polarizations. At the resonance with the lower phonon mode we observed a frequency splitting which points to the excitation of a hybridized state, i.e. a magnon polaron. At resonance with the upper phonon mode we observe a strong driving of the magnon mode by the phonon mode without detecting any frequency splitting. We explain the experimental observations by analyzing the calculated spatial profiles of the phonon modes and show the crucial roles of the phonon mode symmetry and the initial excitation conditions of the magnon polaron by the optical pump pulse.

The direct observation of magnon polarons in the time domain paves the way for exploiting new degrees of freedom for spin excitation and manipulation of spins and phonons at the nanoscale. Magnon polarons allow one to control spins via phonons and phonons via spins when no efficient direct excitation of these modes can be achieved by microwaves or optical techniques. This is explicitly demonstrated in our study, where we gain access to the QTA* phonon mode, while it cannot be excited optically. Moreover, the selectivity in the coupling of magnon and phonon modes allows us to isolate the resulting hybridized state from the optically active phonon mode of the same frequency. This selectivity, which is determined by the spatial overlap of the interacting modes, can be achieved also by alternative ways of spatial matching. For instance, one could imagine asymmetric periodic structures or two-dimensional patterns, the spatial profile of which provides efficient coupling of particular phonon and magnon modes, while other modes remain uncoupled.

The suggested approach is applicable for the manipulation of propagating magnon polarons. While the ability of surface acoustic waves to carry magnons across extremely long distances has been recently demonstrated in the weak coupling regime [44], the strong magnon-phonon coupling will enrich collective magnon-phonon transport by pronounced nonreciprocity and give access to the manipulation of the polarization of surface acoustic waves [42]. Another possibility is the localization of magnon polarons on a defect in a periodic structure due to the gap in the magnon spectrum. This is particularly important in the context of spin-spin interactions and the related phenomenon of magnon Bose-Einstein condensation [3]. An especially appealing prospect is to use the long coherence times and tunability of magnon polarons for creating states with well-defined magnon polaron numbers [45] or coherent magnon-photon-phonon states [46,47], thus providing new routes for quantum information and metrology.

**ACKNOWLEDGEMENTS**

We acknowledge Boris Glavin (†) for his valuable contribution to this work. We are thankful to Mikhail Glazov, Tetiana Linnik and Vitaly Gusev for fruitful discussions. The work was supported by the Bundesministerium fur Bildung und Forschung through the project VIP+ "Nanomagnetron" and by the Deutsche Forschungsgemeinschaft and the Russian Foundation for Basic Research (Grants No. 19-52-12065 and 19-52-12038) in the frame of the International Collaborative Research Center TRR 160. The growth and characterization of the Galfenol-based nanostructures was supported by the Engineering and Physical Sciences Research Council [grant number: EP/H003487/1]. A.N.P. acknowledges for the partial financial support the Russian President Grant No. MD-5791.2018.2.

**References**

[1] M. Weiler, H. Huebl, F. S. Goerg, F. D. Czeschka, R. Gross, S. T. B. Goennenwein, Spin pumping with coherent elastic waves. Phys. Rev. Lett. **108**, 176601 (2012).

[2] T. Kikkawa, K. Shen, B. Flebus, R. A. Duine, K. Uchida, Z. Qiu, G. E. W. Bauer, E. Saitoh, Magnon polarons in the spin Seebeck Effect. Phys. Rev. Lett. **117**, 207203 (2016).

[3] D. A. Bozhko, P. Clausen, G. A. Melkov, V. S. L'vov, A. Pomyalov, V. I. Vasyuchka, A. V. Chumak, B. Hillebrands, A. A. Serga, Bottleneck accumulation of hybrid magnetoelastic bosons. Phys. Rev. Lett. **118**, 237201 (2017).

[4] L. J. Cornelissen, K. Oyanagi, T. Kikkawa, Z. Qiu, T. Kuschel, G. E. W. Bauer, B. J. van Wees, and E. Saitoh, Nonlocal magnon-polaron transport in yttrium iron garnet. Phys. Rev. B **96**, 104441 (2017).

[5] J. Holanda, D. S. Maior, A. Azevedo, S. M. Rezende, Detecting the phonon spin in magnon–phonon conversion experiments. Nature Physics **14**, 500 (2018).

[6] H. Hayashi, K. Ando, Spin pumping driven by magnon polarons. Phys. Rev. Lett. **121**, 237202 (2018).

[7] C. Kittel, Interaction of spin waves and ultrasonic waves in ferromagnetic crystals. Phys. Rev. **110**, 836 (1958).

[8] A. I. Akhiezer, V. G. Bar'iakhtar, and S. V. Peletminskii, Coupled magnetoelastic waves in ferromagnetic media and ferroacoustic resonance. Sov. Phys. JETP **35**, 157 (1959).

[9] J. W. Tucker and V. Rampton, *Microwave ultrasonics in solid state physics* (North-Holland Pub. Co, Amsterdam 1972). pp. 134-183.

[10] O. Yu. Belyaeva, S. N. Karpachev, and L. K. Zarembo, Magnetoacoustics of ferrites and magnetoacoustic resonance. Sov. Phys. Uspekhi **35**, 106 (1992).

[11] A. V. Scherbakov, A. S. Salasyuk, A. V. Akimov, X. Liu, M. Bombeck, C. Bruggemann, D. R. Yakovlev, V. F. Sapega, J. K. Furdyna, and M. Bayer, Coherent magnetization precession in ferromagnetic (Ga,Mn)As induced by picosecond acoustic pulses. Phys. Rev. Lett. **105**, 117204 (2010).

[12] M. Weiler, L. Dreher, C. Heeg, H. Huebl, R. Gross, M. S. Brandt, and S. T. B. Goennenwein, Elastically driven ferromagnetic resonance in nickel thin films. Phys. Rev. Lett. **106**, 117601 (2011).

[13] J.-W. Kim, M. Vomir, and J.-Y. Bigot, Ultrafast magnetoacoustics in nickel films. Phys. Rev. Lett. **109**, 166601 (2012).

[14] D. Afanasiev, I. Razdolski, K. M. Skibinsky, D. Bolotin, S. V. Yagupov, M. B. Strugatsky, A. Kirilyuk, Th. Rasing, and A. V. Kimel, Laser excitation of lattice-driven anharmonic magnetization dynamics in dielectric $FeBO_3$. Phys. Rev. Lett. **112**, 147403 (2014).

[15] Y. Yahagi, B. Harteneck, S. Cabrini, and H. Schmidt, Controlling nanomagnet magnetization dynamics via magnetoelastic coupling. Phys. Rev. B **90**, 140405(R) (2014).

[16] L. Thevenard, C. Gourdon, J. Y. Prieur, H. J. von Bardeleben, S. Vincent, L. Becerra, L. Largeau, and J.-Y. Duquesne, Surface-acoustic-wave-driven ferromagnetic resonance in (Ga,Mn) (As,P) epilayers. Phys. Rev. B. **90**, 094401 (2014).

[17] J. Janušonis, C. L. Chang, P. H. M. van Loosdrecht, and R. I. Tobey, Frequency tunable surface magneto elastic waves. Appl. Phys. Lett. **106**, 181601 (2015).

[18] J. V. Jäger, A. V. Scherbakov, B. A. Glavin, A. S. Salasyuk, R. P. Campion, A. W. Rushforth, D. R. Yakovlev, A. V. Akimov, and M. Bayer, Resonant driving of magnetization precession in a ferromagnetic layer by coherent monochromatic phonons. Phys. Rev. B **92**, 020404(R) (2015).

[19] M. Deb, E. Popova, M. Hehn, N. Keller, S. Mangin, and G. Malinowski, Picosecond acoustic-excitation-driven ultrafast magnetization dynamics in dielectric Bi-substituted yttrium iron garnet. Phys. Rev. B **98**, 174407 (2018).

[20] Y. Hashimoto, D. Bossini, T. H. Johansen, E. Saitoh, A. Kirilyuk, and T. Rasinget, Frequency and wavenumber selective excitation of spin waves through coherent energy transfer from elastic waves. Phys. Rev. B **97**, 140404(R) (2018).

[21] L. Novotny, Strong coupling, energy splitting, and level crossings: a classical perspective. Am. J. Phys. **78**, 1199 (2010).

[22] S. R.-K. Rodriguez, Classical and quantum distinctions between weak and strong coupling. Eur. J. Phys. **37**, 025802 (2016).

[23] C. Berk, M. Jaris, W. Yang, S. Dhuey, S. Cabrini, and H. Schmidt, Strongly coupled magnon–phonon dynamics in a single nanomagnet. Nat. Commun. **10**, 2652 (2019).

[24] J. Atulasimha, and A. B. A. Flatau, A review of magnetostrictive irongallium alloys. Smart Mater. Struct. **20**, 043001 (2011).

[25] D. E. Parkes, L. R. Shelford, P. Wadley, V. Holý, M. Wang, A. T. Hindmarch, G. van der Laan, R. P. Campion, K. W. Edmonds, S. A. Cavill, and A. W. Rushforth, Magnetostrictive thin films for microwave spintronics, Sci. Reports **3**, 2220 (2013).

[26] A. V. Scherbakov, A. P. Danilov, F. Godejohann, T. L. Linnik, B. A. Glavin, L. A. Shelukhin, D. P. Pattnaik, M. Wang, A. W. Rushforth, D. R. Yakovlev, A. V. Akimov, and M. Bayer, Optical excitation of single- and multimode magnetization precession in Fe-Ga nanolayers. Phys. Rev. Applied **11**, 031003 (2019).

[27] A. Bartelsa, R. Cerna, C. Kistner, A. Thoma, F. Hudert, C. Janke, and T. Dekorsy, Ultrafast time-domain spectroscopy based on high-speed asynchronous optical sampling. Rev. Sci. Instrum. **78**, 035107 (2007).

[28] W. K. Hiebert, A. Stankiewicz, and M. R. Freeman, Direct observation of magnetic relaxation in a small permalloy disk by time-resolved scanning Kerr microscopy, Phys. Rev. Lett. **79**, 1134 (1997).

[29] C. Thomsen, H. T. Grahn, H. J. Maris, and J. Tauc, Surface generation and detection of phonons by picosecond light pulses. Phys. Rev. B **34**, 4129-4138 (1986).

[30] D. Nardi, M. Travagliati, M. E. Siemens, Q. Li, M. M. Murnane, H. C. Kapteyn, G. Ferrini, F. Parmigiani, and F. Banfi, Probing thermomechanics at the nanoscale: Impulsively excited pseudosurface acoustic waves in hypersonic phononic crystals. Nano Lett. **11**, 4126–4133 (2011).

[31] See Supplemental Material at https://drive.google.com/file/d/1i81_zww0uVNmhIOzm-tM_s4vvM7vf0Pq/view?usp=sharing for details on the excitation and detection of the NG phonon modes, magnon-phonon interaction in bulk Gaflenol and the model of three coupled harmonic oscillators.

[32] A. S. Salasyuk, A. V. Rudkovskaya, A. P. Danilov, B. A. Glavin, S. M. Kukhtaruk, M. Wang, A. W. Rushforth, P. A. Nekludova, S. V. Sokolov, A. A. Elistratov, D. R. Yakovlev, M. Bayer, A. V. Akimov, and A. V. Scherbakov, Generation of a localized microwave magnetic field by coherent phonons in a ferromagnetic nanograting. Phys. Rev. B **97**, 060404(R) (2018).

[33] M. Langer, R. A. Gallardo, T. Schneider, S. Stienen, A. Roldán-Molina, Y. Yuan, K. Lenz, J. Lindner, P. Landeros, and J. Fassbender, Spin-wave modes in transition from a thin film to a full magnonic crystal. Phys. Rev. B **99**, 024426 (2019).

[34] M. van Kampen, C. Jozsa, J. T. Kohlhepp, P. LeClair, L. Lagae, W. J. M. de Jonge, and B. Koopmans, All-optical probe of coherent spin waves, Phys. Rev. Lett. **88**, 227201 (2002).

[35] V. N. Kats, T. L. Linnik, A. S. Salasyuk, A. W. Rushforth, M. Wang, P. Wadley, A. V. Akimov, S. A. Cavill, V. Holy, A. M. Kalashnikova, A. V. Scherbakov, Ultrafast changes of magnetic anisotropy

driven by laser-generated coherent and noncoherent phonons in metallic films. Phys. Rev. B **93**, 214422 (2016).

[36] R. Verba, I. Lisenkov, I. Krivorotov, V. Tiberkevich, and A. Slavin, Nonreciprocal surface acoustic waves in multilayers with magnetoelastic and interfacial Dzyaloshinskii-Moriya interactions. Phys. Rev. Applied **9**, 064014 (2018).

[37] For pure longitudinal and transverse acoustic waves in the bulk, $\beta_1$ and $\beta_2$ are determined by the respective magneto-elastic coefficients, $b_1$ and $b_2$, and may be obtained analytically (Section 2 of the Supplemental Material [31]). However, for the phonon modes of mixed polarizations localized in the NG, $\beta_1$ and $\beta_2$ are determined by the combinations of $b_1$ and $b_2$ and can be calculated by means of numerical simulation. Thus, the following analysis is based on the values estimated from the experiment.

[38] J. Sadhu, J. H. Lee, and S. Sinha, Frequency shift and attenuation of hypersonic surface acoustic phonons under metallic gratings. Appl. Phys. Lett. **97**, 133106 (2010).

[39] As it has been shown in Ref. [26] the magnon spectrum of a plain Galfenol film grown on a GaAs substrate and covered by a Cr layer, corresponds to the pinning boundary conditions for both interfaces. However, it is reasonable to suggest that in a patterned structure with the Cr cap removed (within and around the grooves), the open surface is characterized by a free boundary condition for the precessing magnetization.

[40] P. F. Herskind, A. Dantan, J. P. Marler, M. Albert, and M. Drewsen, Realization of collective strong coupling with ion Coulomb crystals in an optical cavity. Nature Phys. **5**, 494-498 (2009).

[41] R. C. Le Craw, and H. Matthews, Acoustic wave rotation by magnon-phonon interaction. Phys. Rev. Lett. 8, 397-399 (1962).

[42] R. Q. Scott, and D. L. Mills, Propagation of surface magnetoelastic waves on ferromagnetic crystal substrates. *Phys. Rev. B* **15**, 3545-3557 (1977).

[43] K. Postava, H. Jaffres, A.Schuhl, F.N. Van Dau, M. Goiran, and A.R. Fert, Linear and quadratic magneto-optical measurements of the spin reorientation in epitaxial Fe films on MgO. *J. Mag. Magn. Mater*. **172**, 199 (1997).

[44] B. Casals, N. Statuto, M. Foerster, A. Hernández-Mínguez, R. Cichelero, P. Manshausen, A. Mandziak, L. Aballe, J. M. Hernàndez, and F. Macià, Generation and imaging of magnetoacoustic waves over millimeter distances. Phys. Rev. Lett. **124**, 137202 (2020).

[45] Y. Chu, P. Kharel, T. Yoon, L. Frunzio, P. T. Rakich, and R. J. Schoelkopf, Creation and control of multi-phonon Fock states in a bulk acoustic-wave resonator. Nature **563**, 666 (2018).

[46] X. Zhang, C.-L. Zou, L. Jiang, and H. X. Tang, Cavity magnomechanics, Sci. Adv. **2**. e1501286 (2016).

[47] J. Li, S.-Y. Zhu, G. S. Agarwal, Magnon-photon-phonon entanglement in cavity magnomechanics. Phys. Rev. Lett. **121**, 203601 (2018).

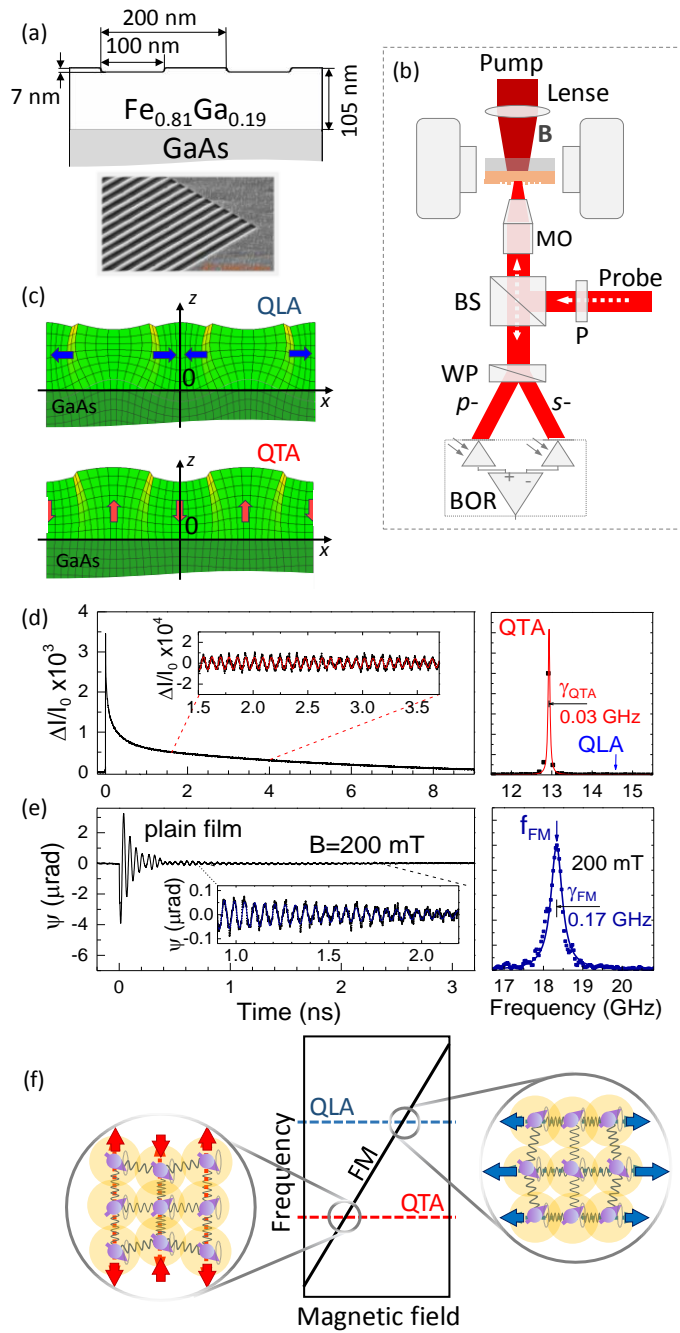

Figure1. Phonons and magnons in a Galfenol nanograting. (a) Design of the studied sample and its SEM image. (b) Experimental scheme: MO – microscope objective, WP – Wollaston prism, BOR – balanced optical receiver, BS – beam splitter, P - polariser. (c) Calculated displacements for two phonon modes excited by the pump pulse. Black arrows illustrate the used coordinate system. The blue and red arrows indicate the main lattice motions for the two modes. (d) Transient reflectivity signal and its fast Fourier transform (FFT) recorded on the grating at conditions when the interaction between the phonon and magnon modes of the NG is fully suppressed (see the text). The inset shows the close-up of the transient signal after subtraction of the slow background. (e) Transient Kerr rotation (KR) signal measured from an unpatterned part of the studied film and its FFT obtained for t>1 ns. The inset shows the zoomed fragment of the transient KR signal, where the fundamental magnon mode solely contributes to the magnetization precession. In both (d) and (e) the transient signals in the insets and their FFTs are shown by symbols, while the fits and their FFT are shown by solid lines. (f) A diagram, which demonstrates the idea of the experiment: the field-dependent magnon mode (oblique line) is tuned into resonance with the field-independent phonon modes (horizontal lines) by an external magnetic field. The crossing points correspond to the FM-QTA and FM-QLA resonance conditions where the magnon-phonon hybridization is expected.

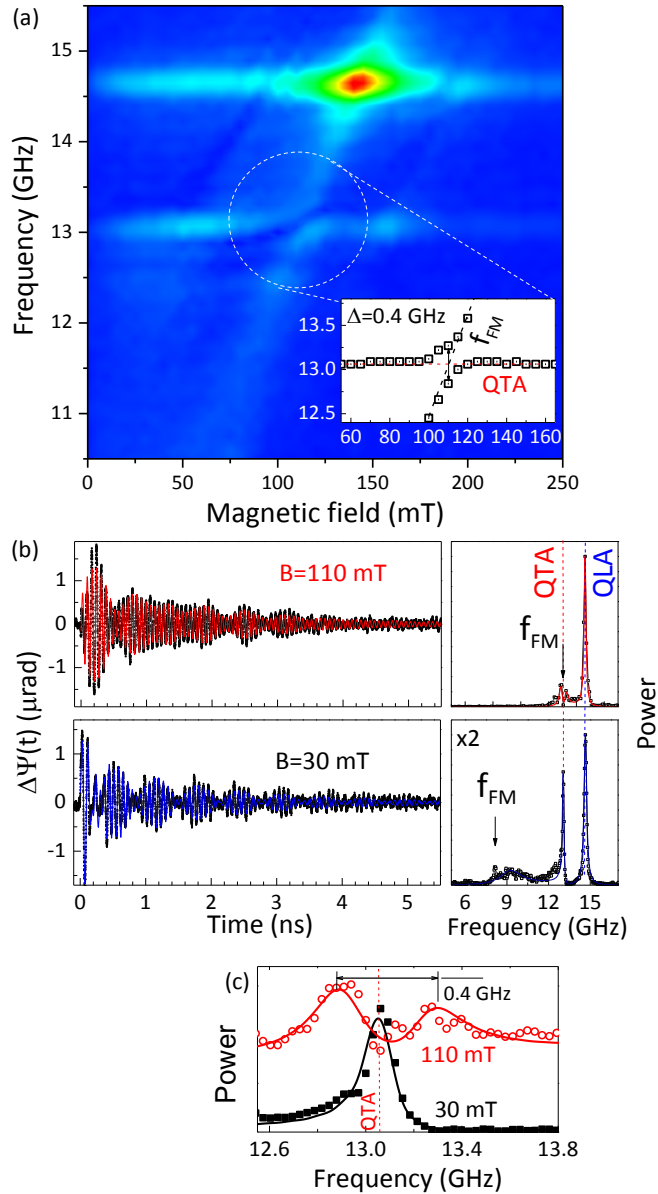

Figure 2. Hybridization of magnon and phonon modes. (a) Color map which shows the spectral density of the measured KR signal as a function of the external magnetic field applied along the NG diagonal when the interaction between the magnon and phonon modes of NG has maximal strength [32]. The anti-crossing is observed at f=13 GHz and B=110 mT. The inset shows the magnetic field dependence of the spectral peaks in the magnon spectrum around the intersection of the QTA and FM modes. (b) Transient KR signals (left panels) and their FFTs (right panels) at non-resonant (B=30 mT) and resonant (B=110 mT) conditions. Symbols show the measured signals and their FFTs; solid lines show respective fits and their FFTs. (c) Zoomed fragments of the FFT spectra shown in (b) around the resonance frequency. The splitting of the line in the resonance at B=110 mT is clearly seen.

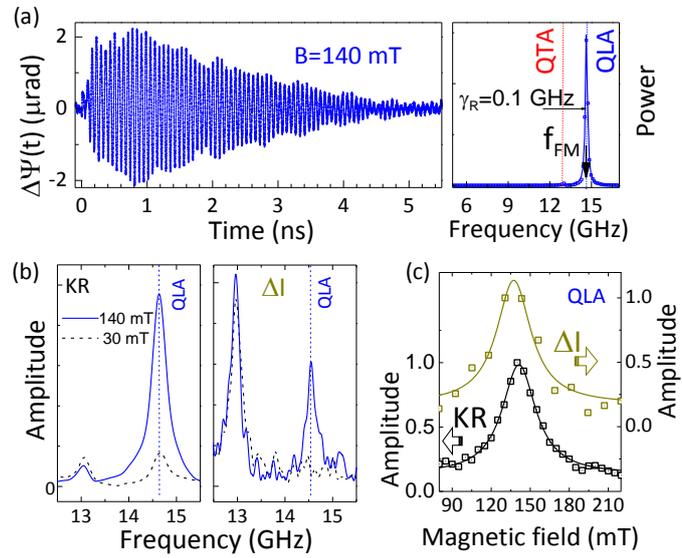

Figure3. Phonon driving of magnons. (**a**) Kerr rotation signal and its FFT measured at *B*=140 mT corresponding to the resonance of the FM and QLA phonon modes. (b) FFTs around $f_{QLA}$ obtained from the KR (left panel) and reflectivity (right panel) signals measured at resonant (*B*=140 mT) and non-resonant (*B*=30 mT) conditions. (c) Magnetic field dependences of the normalized amplitudes at $f=f_{QLA}$=14.5 GHz for the Kerr rotation (lower) and reflectivity (upper) signals (lines are guides for the eye).

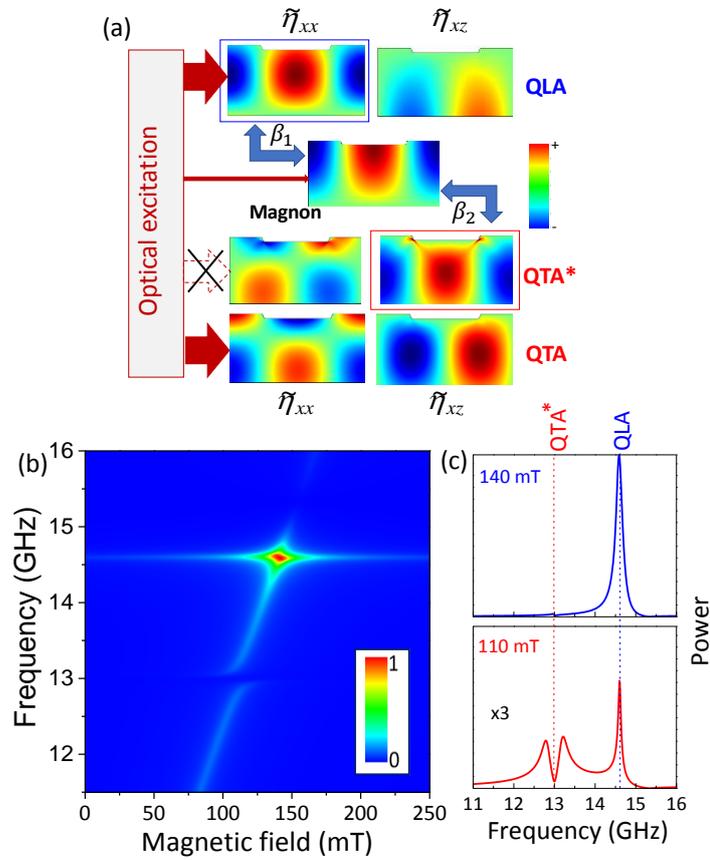

Figure 4. Modeling of the magnon polaron spectrum. (a) Spatial distributions of the normalized dynamical magnetization of the fundamental magnon mode and the normalized strain components for the symmetric QTA, antisymmetric QTA* and symmetric QLA modes. The blue arrows show the pairs, for which the overlap integrals have non-negligible values. (b) The colour map shows the calculated magnetic field dependence of the spectral density of the magnon states calculated in the model of three coupled harmonic oscillators. The initial conditions for the calculations are set as following: the lower QTA* mode is not excited (as mentioned in the text), the energy transfer from the pump pulse to the upper QLA mode significantly exceeds the energy transferred to the FM mode. (c) Calculated magnon spectral power density for magnetic fields corresponding to the resonance of the FM mode with the QTA* mode (B=110 mT) and with the QLA mode (B=140 mT).

18